\documentclass[showpacs,aps,prl,amsmath,amssymb,twocolumn]{revtex4}
\usepackage{graphicx}
\usepackage{dcolumn}
\usepackage{bm,color}
\usepackage{txfonts}
\usepackage{amsmath,epsfig}
\usepackage{epstopdf}
\newcommand{\ket}[1]{\left\vert#1\right\rangle}
\newcommand{\bra}[1]{\left\langle#1\right\vert}

\newcommand{\eq}{Eq.~}
\newcommand{\eqs}{Eqs.~}
\newcommand{\fig}{Fig.~}
\newcommand{\figs}{Figs.~}
\newcommand{\cf} {cf.~}
\newcommand{\ug} {\!=\!}

\newcommand{\piu} {\!+\!}
\newcommand{\meno} {\!-\!}

\newcommand{\eg} {e.g.~}

\newcommand{\rref} {Ref.~}
\newcommand{\rrefs} {Refs.~}

\begin{document}

\author{F. Ciccarello$^1$,  G. M. Palma$^2$, and V. Giovannetti$^1$}
\affiliation{$^1$NEST, Scuola Normale Superiore and Istituto Nanoscienze-CNR, Piazza dei Cavalieri 7, I-56126 Pisa, Italy\\
$^2$NEST, Istituto Nanoscienze-CNR and Dipartimento di Fisica e Chimica, Universit$\grave{a}$  degli Studi di Palermo, via Archirafi 36, I-90123 Palermo, Italy}

% 03.65.Yz Decoherence; open systems; quantum statistical methods
%03.67.-a Quantum information
% 42.50.Lc: Quantum fluctuations, quantum noise, and quantum jumps

\title{Collision-model-based approach to non-Markovian quantum dynamics}
 
\date{\today}
\begin{abstract}
We present a theoretical framework to tackle quantum non-Markovian dynamics based on a microscopic collision model (CM), where the bath consists of a large collection of initially uncorrelated ancillas.
Unlike standard memoryless CMs, we endow the  bath with memory by introducing inter-ancillary collisions between next system-ancilla interactions. Our model interpolates between a fully Markovian dynamics and the continuous interaction of the system with a single ancilla, i.e., a strongly non-Markovian process. We show that in the continuos limit one can derive a general master equation, which while keeping such features is guaranteed to describe an unconditionally completely positive and trace-preserving dynamics. {We apply our theory to an atom in a dissipative cavity for a Lorentzian spectral density of bath modes, a dynamics which can be exactly solved. The predicted evolution shows a significant improvement in approaching the exact solution with respect to two well-known memory-kernel master equations.}
 \end{abstract}
\maketitle

\noindent
%{\it Introduction.}  
{In open system dynamics the focus is on a system ``$S$" in contact with an external  environment. }Typically, the goal is to seek
a master equation (ME) where the degrees of freedom (DOFs) of $S$ are the only explicit variables. Hence, the environmental interactions should be accounted for through an effective but {\it reliable} description. When it comes to quantum objects, this problem turns out to be especially thorny \cite{petruccione,weiss,huelga,breuer}. Within this context, ``reliable" means that the ME to be worked out should give rise to a completely positive and trace-preserving (CPT) dynamics. It is well-assessed that Markovian, i.e., memoryless, environments are described by MEs in the so called Lindblad form \cite{petruccione} entailing unconditionally CPT dynamics. Markovianity is in most cases only an approximation, though: in general, the environment is not forgetful and there is indeed a broad variety of actual phenomena featuring strong {\it non-Markovian} (NM) effects \cite{nm-actual}. Yet, a general systematic framework for describing these has not been developed to date, which is a topical issue of concern to a manifold of variegated research areas. Rather, many different approaches have been proposed \cite{nota-citaz}. Typically, they are to some extent regularly underpinned by phenomenological assumptions and/or approximations (testifying the formidable hurdles to cope with in such problems). As a consequence, non-CPT -- namely unphysical -- dynamics can turn out in certain parameter ranges \cite{vacchini1}. Among these descriptive tools are the so called memory-kernel MEs{, \eg those in \rrefs \cite{barnett, lidar}}. These are integro-differential MEs featuring a history integral, where past states of $S$ are weighted through a certain memory-kernel function. {There exist regimes in which such MEs can fail to be CPT \cite{budini, daffer, sabrina, vacchini2, wilkie, laura, kossa}.  What is more, it was recently tested \cite{laura-breuer} whether MEs in \rrefs\cite{barnett, lidar} are non-Markovian according to a non-Markovianity indicator proposed by Breuer {\it et al.} \cite{measure}. It turned out that this is null \cite{laura-breuer}, which suggests that such MEs should rather be regarded as time-dependent Markovian. This means that when they entail a CPT dynamics this is anyway very close to the purely Markovian regime { \it (weak non-Markovianity)}.}

{In this work we tackle the problem to derive a non-Markovian ME by employing a suitably defined {\it collision model} (CM) \cite{rau,alicki,scarani, buzek, pelle1, pelle2, vittorio,indivisible}
of the system-bath interactions. This allows us to identify a new class of MEs featuring two attractive properties that rarely hold simultaneously. First, they {\it unconditionally} fulfill the CPT condition. Second, they nicely allow to {\it interpolate} between the purely Markovian regime and the {\it strongly non-Markovian} situation where $S$ is continuously  interacting with a 
{low} dimensional, hence non-forgetful, environment. {Also, the model applies regardless of the dimensionality of $S$ and the form of the system-ancilla coupling}.
We recall that in a conventional CM approach  to open dynamics,} the bath is modeled as a large collection of non-interacting identical ancillas (each can be thought as a {low-dimensional system} even though this is not necessary). By hypothesis, $S$ ``collides" with each of these one at a time and, importantly, is not allowed to interact more than once with a given ancilla. Demonstrably, such a process gives rise to an irreversible dynamics for $S$
corresponding to a Lindblad-type \cite{petruccione}, i.e., Markovian, ME \cite{buzek}. This can be expected since, as stressed, at each step $S$ comes into contact with a fresh ancilla which is still in its initial state. Hence, there is no way for the bath to keep track of the system's past history. 
Although they are somewhat fictitious, latest research is unveiling the potential of CMs as effective theoretical tools for tackling open system dynamics \cite{vittorio,indivisible}. First, they are conceptually intuitive, hence potentially easier to cope with: a complex coupling to a large environment is decomposed as a succession of elementary interactions with its subparts. A key feature is that CMs lead to Lindblad-type MEs without demanding any approximation: in fact, only the passage to the continuous limit is needed  \cite{buzek}. This is in contrast to standard microscopic system-reservoir models \cite{petruccione}, where Markovianity must be somehow enforced through drastic assumptions such as the requirement of small coupling and short enough bath correlation time (Born-Markov approximation). Should such a feature be maintained in a NM generalization of a CM, this would be quite appealing: as stressed above, approximations and phenomenological assumptions can lead to unphysical predictions. First progress along this line has been made very recently {\cite{pelle1, vittorio,indivisible}}. {In particular,} Rybar {\it et al.} \cite{indivisible} introduced a CM able to simulate any indivisible channel \cite{petruccione} (thus highly NM) when $S$ is a single qubit \cite{NC}, i.e., a two-level system. Memory was introduced by taking the bath ancillas initially in a nontrivial quantum-correlated state, whose form depends on the specific simulated channel.
{ In the present work we tackle the problem from a completely different perspective: in line with physical intuition, we describe the memory effects  as arising directly from the internal dynamics of the bath itself}. 
Specifically, in the very spirit of standard CMs, a natural memory mechanism to devise is adding inter-ancillary (AA) collisions between next system-ancilla (SA) ones.
This way, quantum information received from $S$ can be conveyed across the bath and returned to $S$ in next SA collisions { (information backflow)}.
Here, we introduce a CM {with memory} precisely built upon this idea. 
In the beginning (see \fig1) $S$ collides with ancilla 1. In standard (Markovian) CMs, $S$-2 collision would then follow, then $S$-3 and so on 
[see \fig1(a)]. This way, each ancilla would still be in the initial state before colliding with $S$, thus fully ``unaware" of previous collisions. In contrast, as sketched in \fig1(b), we assume that an extra AA collision between 1 and 2 occurs after $S$-1 but before $S$-2. Thereby, prior to its interaction with $S$, ancilla 2 will now be in a perturbed state in which information over past history of $S$ is imprinted. The process proceeds by mere iteration: once $S$-2 collision is over, a 2-3 interaction follows, then $S$-3, 3-4 etc. 
\begin{figure}
\includegraphics[width=0.3\textwidth]{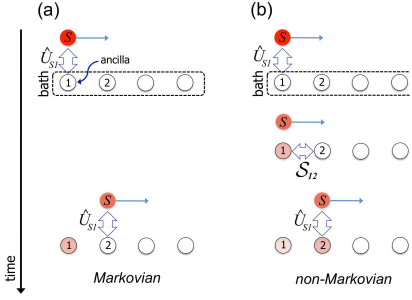}
\caption{(Color online){ {Sketch of the collisional model in the memoryless case (a)  and in the NM one (b). Only the first steps are shown. The next ones are obtained through simple iteration.}}  
 \label{Fig1}}
\end{figure} 
Each collision, either SA or AA, is described by a CPT quantum map affecting the DOFs of the two involved particles. 
 Specifically, {without loss of generality the SA collision involving the $i$th ancilla is defined as the mapping 
  $\sigma \rightarrow \mathcal{U}_{Si}[\sigma]\ug\hat{U}_{Si}\sigma\hat{U}_{Si}^\dag$}, with $\hat{U}_{Si}\ug e^{-i \hat{H}_{Si} \tau}$ being a unitary operator which depends upon the collision time $\tau$ and  the interaction  Hamiltonian $\hat H_{Si}$ (we set $\hbar\ug1$ throughout). 
 Instead, the AA collision involving  the  $i$-th and the $(i+1)$-th ancillas is defined 
 in terms of a stochastic process $\mathcal{S}_{i+1,i}$  which, with probability $p$,  exchanges their states  { or leaves the system unaffected.
 Formally, this is described by the transformation 
\begin{eqnarray}
\sigma \rightarrow \mathcal{S}_{i+1,i}[\sigma]&\ug&  (1-p) \; \sigma\piu p \; \hat{S}_{i+1,i}\sigma\hat{S}_{i+1,i}\,,\label{ipswap}
\end{eqnarray}
where   $\hat{S}_{i+1,i}$ is  the swap operator \cite{NC} on ancillas $i$ and $i+1$. As discussed in the last part of the manuscript, different AA collisional mechanisms can be
selected: the one in~\eq(\ref{ipswap}) however has the advantage that it allows for a simple analytical treatment 
while perfectly capturing the idea of  information backflow mediated by the environment. Here, the parameter $p$ plays the role of a knob for tuning the bath memory.
The overall state of the system at the $n$th step of the evolution is therefore given by 
\begin{equation}\label{evoluzione}
\sigma_{n}=\left( \mathcal{U}_{Sn}\!\circ\!\mathcal{S}_{n,n\meno1}\!\circ \ldots \circ\!\mathcal{U}_{S2}\!\circ\!\mathcal{S}_{2,1}\!\circ\!\mathcal{U}_{S1}\right)[\sigma_0]\;,
\end{equation}
 where ``$\circ$" represents the super-operator composition and will be henceforth omitted and 
$\sigma_0\ug \rho_0 \ket{{\bf 0}}_B\!\bra{{\bf 0}}$ 
  is the system-bath initial state \cite{notanotaz} with  
    $\rho_0$  being the input density matrix of  $S$  and $\ket{{\bf 0}}_B\ug\ket{0}_1\!\ket{0}_2 \cdots$ the initial ancillary state \cite{purification}. Exploiting the properties of the swap operator and the translational symmetry of the environmental initial state, we find it useful to cast  Eq.~(\ref{evoluzione})  in a recursive form, where $\sigma_n$ is expressed as a sum of terms involving 
states $\{\sigma_{m<n}\}$. Specifically, for $n\!\ge\!2$ we straightforwardly obtain \cite{sm}%this is generalized by induction~\cite{NOTA1}  as
\begin{equation}\label{sigman}
\sigma_n\ug (1-p)\sum_{j=1}^{n\meno1}p^{j-1}{\cal U}_{Sn}^j\,[\sigma_{n\meno j}] \; \piu \;  p^{n-1} {\cal U}^n_{Sn}[\sigma_0]\,,
\end{equation}
where now ${\cal U}^{j}_{Sn}$ represents $j$ consecutive applications  of the unitary gate ${\cal U}_{Sn}$, i.e., ${\cal U}_{Sn}^j[\sigma]\ug e^{-i \hat{H}_{Si} j\tau}\sigma e^{i \hat{H}_{Si} j\tau}$. Note that this corresponds to a coherent interaction process between $S$ and the $n$th ancilla {\it only}, {which continued} for a time $j\tau$. 
This and the fact that in \eq(\ref{sigman}) each ${\cal U}_{Sn}^j$ is applied to $\sigma_{n-j}$ (with $n$ still in $\ket{0}_n$) entail the attractive property that
an expansion for $\rho_n\ug{\rm Tr}_B\sigma_n$ similar to \eq(\ref{sigman}) holds. Tracing this over $B$ indeed yields 
\begin{equation}\label{rhon}
\rho_n\ug (1-p)  \sum_{j=1}^{n\meno1}p^{j-1}\mathcal{E}_j[\rho_{n\meno j}]\piu  p^{n-1} \mathcal{E}_n[\rho_0]\,,
\end{equation}
where a transformation $\mathcal{E}_j$ is a  CPT map on $S$ {\it only} defined in terms of the unitary map ${\cal U}_{Sn}^j$ and the initial bath state as
\begin{eqnarray}\label{map}
\mathcal{E}_j[\rho] \ug \mbox{Tr}_B \left\{  {\cal U}_{Sn}^j [ \rho \otimes |\bf {0}\rangle_{B}\langle \bf{0}|] \right\} \;.
\end{eqnarray}}
Interestingly, the structure of \eq(\ref{rhon}) shares features with the discrete model used by Shabani and Lidar \cite{lidar} to derive their ME { (there, in particular, $\mathcal{E}_j$ is the dynamical map in absence of measurements performed on the bath).}  
Two major differences occur, though. First, \eq(\ref{rhon}) cannot be written as a single sum due to the missing $(1\meno p)$ factor in the last term, which in fact means that here we deal with a time-{\it inhomogeneous} memory-kernel function (MKF). Second, map $\mathcal{E}_j$ is in general strongly NM: { it describes the reduced dynamics of $S$ for a continuous coherent interaction between $S$ and a {\it single} ancilla (\eg once can think of two coupled spins periodically exchanging an excitation)}.  Indeed, {  as anticipated earlier,} our model interpolates between two extreme regimes depending on the value of the probability $p$. When $p\ug0$, AA collisions are absent [\cf\eq(\ref{ipswap})]: \eq(\ref{rhon}) reduces to $\rho_n\ug\mathcal{E}_1[\rho_{n-1}]$ and we retrieve a  standard Markovian CM \cite{rau,alicki,scarani, buzek}.
Quite differently, for $p\ug1$, \eq(\ref{rhon}) yields $\rho_n\ug \mathcal{E}_n[\rho_{0}]$, i.e., $S$ behaves as if it interacts with a single ancilla all the time. This can be seen by noting that for $p\ug1$ \eq(\ref{ipswap}) reduces to a perfect swap: once $S$ has undergone a $\tau$-long interaction with $i$, the final state of $i$ is fully transferred to $i\piu1$ (with $i$ returning to $\ket{0}_i$).

Our next goal is to work out the ME corresponding to \eq(\ref{rhon}) in the continuous limit and then prove
{that (i) the resulting equation is still capable to interpolate between the two opposite limits depicted above and (ii) it unconditionally satisfies the CPT condition.
For this aim, we first subtract from \eq(\ref{rhon}) the analogous identity for $n\meno1$.  This gives rise to an equation for the variation of $\rho_n$ between two next steps $\Delta \rho_n\ug\rho_n\meno\rho_{n\meno1}$, which reads  
\begin{eqnarray}\label{fd}
\!\Delta \rho_n&\ug& (1\meno p) \!\sum_{j=1}^{n\meno2}p^{j-1}\mathcal{E}_j[\Delta \rho_{n\meno j}]\piu (1\meno p)p^{n\meno1} \mathcal{E}_{n\meno1}[\rho_1]
\nonumber \\
&&\piu \;\; \Delta \left(p^{n\meno1} \mathcal{E}_n\right)[\rho_0].\!\!\!\!\,\,\,\,\,\,\,
\end{eqnarray}
This can now be transformed into a differential equation for the continuous time evolution of the system density matrix $\rho(t)$ 
by taking the limit of infinite collisions [$n,j \rightarrow \infty$] while sending the
collision time to zero [i.e. $\tau\rightarrow 0$] in such a way that the elapsed times $t=n\tau$  and $t'= j \tau$ remain finite.  
Also, when $j$ becomes very large the probability $p^j$ of multiple AA collisions clearly must not vanish.
We thus set $p = \exp[-\Gamma \tau]$, where $\Gamma\ug-(\log p)/\tau$ is interpreted as the {\it memory rate}. We require that, when $\tau\!\rightarrow\!0$, $p$ {approaches} 1 in such a way that $\Gamma$ remains finite. This allows to express each power of $p$ as a decaying exponential $p^{j}\ug ({p^{\frac{1}{\tau}}})^{j\tau}\ug e^{-\Gamma t'}$. Note that in the continuous limit $\tau$ should be far shorter than any characteristic time, in particular $\Gamma^{-1}$. This gives $\Gamma\tau\!\ll\!1$ and thus $1\meno p\ug1\meno e^{-\Gamma \tau}\!\simeq\!\Gamma \tau$. Using this, the sum over $j$ in \eq(\ref{fd}) becomes a time integral {as $\tau\!\rightarrow\!0$}. By identifying $\Delta \rho_n/\tau\!\rightarrow\!\dot{\rho}(t)= \!d\rho(t)/dt$, after a few straightforward steps \cite{sm} we end up with the ME
\begin{equation}\label{ME}
\dot{\rho}(t) =\Gamma\!\int_{0}^t \!\!dt'e^{-\Gamma t'} \;\mathcal{E}({t'})\left[ \dot{\rho}(t-t') 
 \right]+e^{-\Gamma t}\; \dot{\mathcal{E}}(t) [\rho_0]\;,
\end{equation}
where the CPT map $\mathcal{E}(t)$ is the continuous analogue of \eq(\ref{map}) and {the dot stands for the total derivative.}} This is an integro-differential equation in $\rho(t)$ featuring a history integral term with an associated MKF $\Gamma e^{-\Gamma t'}$ and, notably, a term $\sim\!\!\rho_0$. The latter, which stems from the formerly discussed inhomogeneity of the discrete MKF in \eq(\ref{rhon}), is a strong signature of NM behavior. Indeed, {in the limit where memory effects persist  indefinitely, i.e., $\Gamma\!\rightarrow\!0$, it is the only term surviving in \eq(\ref{ME}) yielding $\dot{\rho}(t)\!\rightarrow\!\dot{\mathcal{E}}(t)[\rho_0]$, i.e., $\rho(t)\!\rightarrow\!\mathcal{E}(t)[\rho_0]$} { in full analogy with the discrete model (we address the opposite limit $\Gamma\!\rightarrow\!\infty$ later on)}. Next, we derive the solution of \eq(\ref{ME}) $\rho(t)\ug\Lambda(t)[\rho_0]$ and prove that the dynamical map \cite{petruccione} $\Lambda(t)$ is {\it always} CPT [$\Lambda(0)\ug\mathcal{I}$ with $\mathcal{I}$ the identity superoperator]. Evidently, $\Lambda(t)$ obeys \eq(\ref{ME}) under the formal replacement $\rho\!\rightarrow\!\Lambda$. By taking the Laplace transform (LT) of such equation, this is easily solved as \cite{sm} 
 \begin{eqnarray} \label{ffd}
 \tilde{{\Lambda}}(s)=    \frac{ \tilde{\mathcal E}(s+\Gamma)}{{\mathcal I} - \Gamma\;\tilde{\mathcal E}(s+\Gamma)}\;,\,\,
\end{eqnarray} 
where $\tilde{\Lambda}(s)$ and $\tilde{\mathcal{E}}(s)$ are the LTs of $\Lambda(t)$ and $\mathcal{E}(t)$, respectively [\eq(\ref{ffd}) is well-defined since the numerator and denominator commute].
Expanding \eq(\ref{ffd}) in powers of $\Gamma$ gives $ \tilde{{\Lambda}}(s)\ug\sum_{k=1}^\infty\left[\tilde{\mathcal{E}}(s\!+\!\Gamma)\right]^k\!\Gamma^{k\!-\!1}$, whose inverse LT is
  \begin{eqnarray} \label{ffd3}
\Lambda(t) \ug\mathcal{L}^{-1}[\tilde{\Lambda}(s)](t)\,\ug  \sum_{k=1}^\infty\Gamma^{k\!-\!1} \; \mathcal{L}^{-1}\![\tilde{\mathcal{E}}^k(s\!+\!\Gamma)](t)\;.
\end{eqnarray} 
Basic properties of LT allow to immediately calculate the inverse LT within braces as \cite{sm}
  \begin{eqnarray} 
\!\mathcal{L}^{\meno1}\![\tilde{\mathcal{E}}^k\!(\!s\!+\!\Gamma\!)] &\ug& e^{-\Gamma t}\!\!\!\int_0^t \!\!\!dt_1\!\!\!\int_0^{t_1}\!\!\!\!\!\!dt_2\!\cdot\!\cdot\!\cdot\!\!\!\!\int_0^{t_{k\meno2}} \!\!\!\!\!\!\!d t_{k\meno1} \nonumber \\
&&\times \; \mathcal{E}(t_{k\meno1}\!) \mathcal{E}(t_{k\meno2}\!\meno \!t_{k\meno1}\!)\cdot\!\cdot\!\cdot\!\mathcal{E}(t\!\meno\! t_1\!).\,\,\, \,
\label{ltn}
\end{eqnarray} 
We have thus expressed $\Lambda(t)$ as a weighted series of multiple auto-convolutions of the CPT map $\mathcal{E}(t)$. { The {\it integrand} in \eq(\ref{ltn}) is evidently a composition of CPT $\mathcal{E}$ maps, hence it is CPT itself. Therefore, we see that the dynamical map \eq(\ref{ffd3}) is in fact a combination of CPT maps with {\it positive} weights [factors $\Gamma^{k-1}$ and $e^{-\Gamma t}$ in \eqs(\ref{ffd3}) and (\ref{ltn}) are all positive]. This proves the complete positivity of map $\Lambda(t)$.} Moreover, the state obtained by applying the {\it integrand} in \eq(\ref{ltn}) {(a CPT map as discussed)} to $\rho_0$ has evidently unitary trace. As is easily checked \cite{sm}, this entails ${\rm Tr}\left\{\Lambda(t)[\rho_0]\right\}\ug1$. We conclude that{, since $\mathcal{E}(t)$ is CPT, }
$\Lambda(t)$ is CPT. {The last remaining task is to prove that,} in line with \eq(\ref{rhon}) for $p\ug0$, the Markovian behavior arises from \eq(\ref{ME}) for $\Gamma\!\rightarrow\!\infty$. Indeed, \eq(\ref{ME}) is such that for $\Gamma$ large enough we can approximate ${\mathcal E}(t) \!\simeq\! {\mathcal I} \piu {\mathcal F}t$, where $\mathcal{F}\ug\dot{\mathcal{E}}(0)$. Under LT, this becomes $\tilde{\mathcal{E}}(s)\ug{1}/{(s\piu \Gamma)} \piu {{\mathcal F}}/{ (s\piu\Gamma)^2}$, which once plugged into \eq(\ref{ffd}) and in the limit of $\Gamma\!\rightarrow \!\infty$ yields  $\tilde{{\Lambda}}(s)\ug \left. {(s\piu \Gamma \piu {\mathcal F} )}/{[s^2 \piu \Gamma(s\meno{\mathcal F})]} \right|_{\Gamma \rightarrow \infty}\ug{1}/{(s\meno {\mathcal F})}$. By transforming back, we end up with $\Lambda(t)\ug e^{\mathcal{F}t}$ entailing that the semigroup property is fulfilled and thus, necessarily, $\mathcal{F}$ is a Lindbladian superoperator \cite{petruccione} with \eq(\ref{ME}) reducing to the Lindblad-form $\dot{\rho}(t)\ug\mathcal{F}[\rho]$.
{It is worth noticing that a Markovian dynamics can also be generated from~(\ref{ME}) for a {\it finite} $\Gamma$, by properly choosing the integral generator map ${\mathcal{E}}(t)$. Indeed, if one assume $\mathcal{E}(t)\ug e^{\mathcal{F}t}$,  then  $\Lambda(t)\ug e^{\mathcal{F}t}$ is the {\it exact} solution of \eq(\ref{ME}) for {\it any}~$\Gamma$: 
in other words, the mapping  $\mathcal{E}(t) \rightarrow \Lambda(t)$ induced by~(\ref{ffd}) is transparent for dynamical semigroups}. {This seems reasonable: no information can propagate through AA collisions if each SA collision is already forgetful.}{ 
This further marks the difference from the Shabani-Lidar ME~\cite{lidar} where the fact that $\Lambda(t)\ug e^{\mathcal{F}t}$ is not a solution in that context is exploited in a perturbative way to deliver the conclusions, and strengthens the importance of the last term in \eq(\ref{ME}). Also note that our ME agrees with the general form predicted by Nakajima and Zwanzig~\cite{petruccione} provided that the corresponding $t\meno t'$-dependent  memory-kernel superoperator exhibits a discontinuity at $t\ug t'$ [through part integration method, \eq (\ref{ME}) can be easily expressed in a form where the time integral involves $\rho(t\meno t')$ instead of $\dot{\rho}(t\meno t')$].}

{To test the predictive power of our approach, we consider the dynamics of a two-level atom [whose ground (excited) state is denoted by $\ket{0}_S$ ($\ket{1}_S$)] coupled to a continuum of electromagnetic modes in the rotating-wave approximation \cite{petruccione}. The case in which the field spectral density $J(\omega)$ is a Lorentzian centered on the atomic frequency can be solved {\it exactly} \cite{garraway}, which makes it a useful benchmark to assess the effectiveness of a ME \cite{sabrina}. 
This solution can be expressed in terms of an amplitude damping channel (ADC) \cite{NC} as $\rho(t)\ug \mathcal{A}_{G(t)}[\rho_0]$, where $\mathcal{A}_\eta[\rho_0]\ug1\meno p \vert \eta\vert^2\ket{0}_S\!\bra{0}\piu p \vert \eta\vert^2 \ket{1}_S\!\bra{1}\piu \left\{r \,\eta \ket{0}_S\!\bra{1}\piu{\rm H.c.}\right\}$ is the general form of an ADC ($p$ and $r$ are the atom's initial populations and coherences). Specifically \cite{petruccione}, $G(t)\ug e^{-\lambda/2t}[\cosh(dt/2)+\lambda/d\sinh (dt/2)]$ with $d\ug\sqrt{\lambda^2\meno2\gamma_0\lambda}$. Here, $\lambda$ measures the width of $J(\omega$), while $\gamma_0$ is related to the strength of the coupling \cite{petruccione}. For $\lambda\!\gg\!\gamma_0$, $J(\omega)$ becomes about flat and $G(t)\!\rightarrow\!e^{-\gamma_0/2t}$: the atom undergoes standard spontaneous emission at a rate $\gamma_0$ and $\dot{\rho}\!\rightarrow\!\mathcal{L}[\rho]$, namely the Markovian regime occurs ($\mathcal{L}$ is the usual zero-temperature atomic Lindbladian \cite{petruccione} with associated rate $\gamma_0$). For $\lambda\!<\!\gamma_0/2$, instead, damped oscillations take place as a signature of non-Markovianity. In particular, in the regime $\lambda\!\ll\!\gamma_0$, $G(t)\!\simeq\!e^{-\lambda t}\!\cos(\Omega t)$ with $\Omega\ug\sqrt{\gamma_0\lambda/2}$ showing that the atom undergoes damped Rabi oscillations at a rate $\Omega$ due to its coupling to the cavity protected mode. For vanishing $\lambda$ (ideal cavity with infinite quality factor) we would thus obtain $\mathcal{A}_{G(t)}[\rho_0]\simeq\mathcal{A}_{\cos(\Omega t)}[\rho_0]$. This strongly suggests to regard the cavity protected mode as a generic ancilla in our CM framework and thus set $\mathcal{E}(t)\!\equiv\!\mathcal{A}_{\cos(\Omega t)}$ and, additionally, $\Gamma\!\equiv\!\lambda$. Indeed, we have shown earlier that  if $\Gamma$, namely $\lambda$, vanishes, the system behaves as if interacting all the time with a single ancilla, namely the protected mode. On the other hand, we have seen that when $\Gamma$ is very large (Markovian limit) at each collision the system interacts with a fresh ancilla still in the initial state. Note that even this case can be viewed as an effective {\it single}-ancilla process if one supposes such ancilla to be reset to its initial state between two next collisions with $S$. Correspondingly, in the atom-field model, for very large $\lambda$ the cavity quality factor is very low: the leakage of the protected mode is so effective that at any time -- not only at the beginning -- the atom in fact ``sees" such mode in its vacuum state. With the above settings [$\mathcal{E}(t)\!\equiv\!\mathcal{A}_{\cos(\Omega t)}$ and $\Gamma\!\equiv\!\lambda$] the dynamical map $\Lambda(t)$ can be calculated exactly through \eqs(\ref{ffd})--(\ref{ltn}). In \fig2, we display the time behavior of the atomic excitation, i.e., the excited-state population, and coherences (normalized to the respective initial values) as given by the exact solution and our CM. For comparison, we also report the corresponding functions predicted by the memory-kernel MEs $\dot{\rho}\ug\mathcal{L}\int \!dt' k(t')\rho(t\meno t')$ \cite{barnett} and $\dot{\rho}\ug\mathcal{L}\int \!dt' k(t')e^{\mathcal{L}t'}\rho(t\meno t')$ \cite{lidar} for  $k(t')\ug\lambda e^{-\lambda t'}$, to which we will refer as phenomenological and Shabani-Lidar MEs, respectively (a similar comparison was carried out in \rref\cite{sabrina}). 
\begin{figure}
\includegraphics[width=0.48\textwidth]{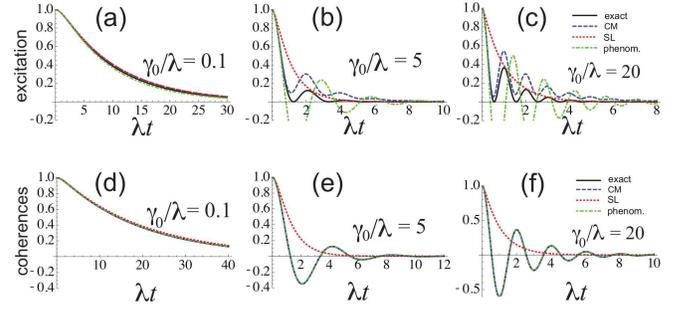}
\caption{(Color online){ { Open dynamics of a two-level atom in contact with a bath of Lorentzian spectral density centered at the atomic frequency. Panels (a)-(c) [(d)-(f)] show the excited state population (coherences) against the rescaled time $\lambda t$ predicted by the exact solution (black solid line), our CM (blue dashed), the Shabani-Lidar ME (red dotted) and the phenomenological ME (green dot-dashed) for different values of $\gamma_0/\lambda$. All the plotted quantities are normalized to the respective initial values.}}  
 \label{Fig1}}
\end{figure} 
For large $\lambda$ (compared to $\gamma_0$) the Markovian regime occurs: all the models basically yield the same purely exponential behavior [see \figs2(a) and (d)]. As $\lambda$ becomes low, significant deviations arise. The Shabani-Lidar model keeps predicting exponential decays [\cf\figs2(b),(c),(e) and (f)] in contrast to the damped oscillations predicted by the exact solution. The phenomenological ME predicts coherences matching the exact solution [\cf\figs2(e) and (f)], however positivity is drastically violated \cite{barnett, sabrina} [see \figs2(b) and (c)]. Our ME \eq(\ref{ME}) yields a substantial improvement on {\it both} the above models. As the phenomenological ME, it accurately reproduces the exact coherences [see \figs2(e) and (f)]. Quite differently, though, it does not break positivity [see \figs2(b) and (c)] in line with our general proof. The latter feature is shared with the Shabani-Lidar ME. Yet, unlike this, the CM captures the physics of the process far better:  damped oscillations for populations rather close to the exact ones are predicted (the discrepancy decreases as $\lambda/\gamma_0\!\rightarrow\!0$). Note that, while in \figs2(b) and (c) the minima of the exact solution are zero, the corresponding CM minima are small but strictly positive. This is most likely to stem from the {\it incoherent} mixture of the identity and swap operator entering \eq(\ref{ipswap}). If this is replaced by a coherent sum (in which case every AA collision becomes unitary) a regime featuring damped oscillations with zero minima indeed occurs \cite{inprep}.

To summarize, we have introduced a NM microscopic CM, where the bath memory is added dynamically through simple inclusion of inter-ancillary collisions each modeled as a CPT swapping operation. The model interpolates between two extreme situations: a fully Markovian regime (retrieving a standard CM) and a strongly NM one (corresponding to the continuous interaction of the system with a single ancilla). The continuous limit gives rise to a ME which was proven to be unconditionally CPT. To test the effectiveness of our approach, we have applied it to an atom coupled to a bath of modes featuring a Lorentzian spectral density and compared the outcomes with the analytical solution and two popular memory-kernel MEs. While all the advantageous features of such MEs {\it simultaneously} occur in ours, this in addition succeeds to capture distinctive traits of the NM dynamics. 

We thank R. Fazio for comments and discussions and  acknowledge support from the MIUR through the FIRB-IDEAS project RBID08B3FM. 

\section*{Supplementary Material}

In this Supplementary Material, we supply some technical details related to the derivation of some properties discussed the paper's main text.

\section{Derivation of \eq(3)}

Map $\mathcal{S}_{i+1,i}$ is given by \eq(1). By definition, $\hat{S}_{i+1,i}$ swaps the states of $i$ and $(i\piu1)$, hence it transforms each $\hat{U}_{Si}$ as $\hat{S}_{i+1,i}\hat{U}_{Si}\,\hat{S}_{i+1,i}\ug \hat{U}_{S,i+1}$ (we recall that $\hat{S}_{ij}^\dag\!\equiv\!\hat{S}_{ij}$). Equivalently, $\hat{S}_{i+1,i}\hat{U}_{Si}\ug \hat{U}_{S,i+1}\,\hat{S}_{i+1,i}$: a swap operator on the left of a $\hat{U}$-type one can jump to the right side provided that the ancillary index of $\hat{U}$ is increased by one unity.  {By construction, 
$\sigma_{2}\ug\mathcal{U}_{S2}\mathcal{S}_{21}[\sigma_1$], i.e., 
$\sigma_2\ug (1\meno p) \hat{U}_{S2}\,\sigma_1\hat{U}_{S2}^\dagger\piu p \hat{U}_{S2}\hat S_{21}(\hat{U}_{S1}\,\sigma_{0}\hat{U}_{S1}^\dagger) \hat S_{21}\hat{U}_{S2}^\dagger$, where we have replaced (only in the latter term) $\sigma_{1}\ug \hat{U}_{S1}\,\sigma_{0}\hat{U}_{S1}^\dagger$ [see \fig1(b)].} Operator $\hat{U}_{S1}$ can be eliminated as follows. We use $\hat{S}_{21}\hat{U}_{S1}\ug \hat{U}_{S,2}\,\hat{S}_{21}$ alongside $\hat{S}_{21}\sigma_0\ug \rho_0 \hat{S}_{21}\!\ket{{\bf 0}}_B\!\bra{{\bf 0}}\!\equiv\!\sigma_0$. This yields
$\sigma_2\ug (1-p) \hat{U}_{S2}\,\sigma_1\hat{U}_{S2}^\dagger\piu p \hat{U}_{S2}^2\,\sigma_0 \,\!\left(\hat{U}_{S2}^\dagger\right)^2$, which indeed corresponds to \eq(3) in the main text for $n\ug2$. 

The arbitrary-$n$ case can be proven by induction as follows. By construction, $\sigma_{n+1}\ug\mathcal{U}_{S,n+1}\mathcal{S}_{n+1,n}[\sigma_n]\ug (1\meno p) \hat{U}_{S,n+1}\,\sigma_n\hat{U}_{S,n+1}^\dagger\piu p \hat{U}_{S,n+1}\hat S_{n+1,n}\sigma_n \hat S_{n+1,n}\,\hat{U}_{S,n+1}^\dagger$. By replacing \eq(3) in the second term and using $\hat{S}_{n+1,n}\hat{U}_{sn}\ug\hat{U}_{s,n+1}\hat{S}_{n+1,n}$ along with $\hat{S}_{n+1,n}\sigma_{n-j}\ug \sigma_{n-j}$ for $j\!\ge\!1$, we end up with \eq(3) for $n\!\rightarrow\!n\piu1$.

\section{Derivation of \eq(7)}

When \eq(6) in MT is divided by $\tau$ and by using the limiting expressions discussed in the main text, the terms on the right-hand side in the continuous limit take the form
\begin{eqnarray}\label{fd2}
&&\frac{c^2 \sum_{j=1}^{n\meno2}s^{2(j-1)}\mathcal{E}_j\left[{\rho_{n\meno j}\meno\rho_{n\meno1\meno j}}\right]}{\tau}\!\simeq\!\Gamma\! \int_{0}^t \!dt' e^{-\Gamma t'}\mathcal{E}(t')\left[\frac{d\rho(t\meno t')}{d (t\meno t')}\right]\,\,\nonumber\\
&&\frac{\Delta (s^{2(n\meno1)} \mathcal{E}_n)}{\tau}\,[\rho_0]\ug\frac{(s^{2(n\meno1)} \mathcal{E}_n\meno s^{2(n\meno2)} \mathcal{E}_{n\meno1})}{\tau}\,[\rho_0]\nonumber\\
&&\,\,\,\,\,\,\,\,\,\,\,\,\,\,\,\,\,\,\,\,\,\,\,\,\,\,\,\,\,\,\,\,\,\,\,\,\!\!\!\!\!\!\!\!\!\!\!\!\!\!\simeq\!\frac{e^{-\Gamma (t+ 2\tau)}\mathcal{E}_{t+\tau}\meno e^{-\Gamma (t+ \tau)}\mathcal{E}(t)}{\tau}\,[\rho_0] \ug \frac{d}{dt}\!\left(e^{-\Gamma t }\mathcal{E}(t)\right)[\rho_0]\,\nonumber\\
 && \frac{c^2s^{2(n\meno1)} \mathcal{E}_{n\meno1}}{\tau}\,[\rho_1]\!\simeq\!\Gamma e^{-\Gamma t}\mathcal{E}(t)\, [\rho_0]\,\,\nonumber
\end{eqnarray}
whereas the left-hand side of \eq(6) in MT clearly reduces to the time derivative of $\rho(t)$. Using these, \eq(7) is immediately obtained.

\section{Derivation of $\tilde{\Lambda}(s)$} \label{sollaplace}

By replacing $\rho(t) = \Lambda(t) [\rho_0]$ in MT's \eq(7) and using that $\rho_0$ is arbitrary, the equation obeyed by $\Lambda(t)$ tuns out to be 
\begin{eqnarray}
\dot{\Lambda}(t) = \Gamma\! \int_0^t dt' \exp[-\Gamma t'] \; {\mathcal E}(t') \dot{\Lambda}(t-t')
+ \exp[-\Gamma t] \;\dot{{\mathcal E}}(t)\,\,\,\,\,\,\,\,
\end{eqnarray} 
in addition to the requirement ${\Lambda}(0)\ug{\mathcal I}$.
Upon Laplace transform (LT), the equation becomes
\begin{eqnarray} \label{ltmes}
s \tilde{{\Lambda}}(s) \meno{\mathcal I}  \ug \Gamma \tilde{\mathcal E}(s+\Gamma)  [s \tilde{{\Lambda}}(s) \meno {\mathcal I} ]  \piu
(s +\Gamma) \; \tilde{\mathcal E}(s+\Gamma) - {\mathcal I}\,\,\,\,\,\,
\end{eqnarray} 
where for $s$ complex the LT is defined as 
\begin{equation}
\tilde{F}(s) \ug \mathcal{L}\,[F(t)](s)\ug \int_0^\infty\!\! d t \; e^{-s t} F(t)\,
\end{equation} 
and we have used $\mathcal{E}(0)\ug\mathcal{I}$ (see main text). 

By rearranging terms in \eq(\ref{ltmes}) 
  \begin{eqnarray} 
[ {\mathcal I} -  \Gamma \; \tilde{\mathcal E}(s+\Gamma) ]  [s \tilde{{\Lambda}}(s) - {\mathcal I}  ]=  
(s +\Gamma) \; \tilde{\mathcal E}(s+\Gamma) - {\mathcal I}\;,
\end{eqnarray} 
and hence 
 \begin{eqnarray} 
s \; [ {\mathcal I} -  \Gamma \; \tilde{\mathcal E}(s+\Gamma) ] \  \tilde{{\Lambda}}(s)=  
s  \; \tilde{\mathcal E}(s+\Gamma) \;.
\end{eqnarray} 
By simplifying $s$ on both terms and introducing the inverse of ${\mathcal I} -  \Gamma \; \tilde{\mathcal E}(s+\Gamma)$ we end up with \eq(8).

\section{Expansion of $\Lambda(t)$}

The inverse LT of $\tilde{\mathcal{E}}(s+\Gamma)$ is $\mathcal{L}^{-1}[\tilde{\mathcal{E}}(s+\Gamma)]\ug e^{-\Gamma t}\mathcal{E}(t)$. Thereby, from a basic property of LT, the inverse transform of $\tilde{\mathcal{E}}^2(s\!+\!\Gamma)$ is the auto-convolution of $e^{-\Gamma t}\mathcal{E}(t)$, which reads
\begin{eqnarray}
\mathcal{L}^{-1}[\tilde{\mathcal{E}}^2(s+\Gamma)]&\ug& \!\int_0^t \!dt' \left[ e^{-\Gamma t'}\mathcal{E}(t') \right]\left[ e^{-\Gamma (t\meno t')}\mathcal{E}(t\meno t')\right]\nonumber\\
&\ug& e^{-\Gamma t}\!\! \int_0^t \!dt' \mathcal{E}(t') \mathcal{E}(t\meno t')\,.\label{lt2}
\end{eqnarray}
The inverse LT of $\tilde{\mathcal{E}}^3(s\!+\!\Gamma)$ can be calculated as the convolution between \eq(\ref{lt2}) and $\mathcal{L}^{-1}[\tilde{\mathcal{E}}(s+\Gamma)]\ug\mathcal{E}(t)$, which yields
\begin{equation}\label{lpt3}
\mathcal{L}^{-1}[\tilde{\mathcal{E}}^3(s\!+\!\Gamma)]\ug e^{-\Gamma t}\!\int_0^t \!dt_1\!\! \int_0^{t_1}\! \!dt_2 \,\mathcal{E}(t_2\!) \mathcal{E}(t_1\!\meno\! t_2)\mathcal{E}(t\meno t_1)\,\,.
\end{equation}
\eq(10) in MT, i.e., the case corresponding to $\tilde{\mathcal{E}}^k(s\piu\Gamma)$ for arbitrary $k$, then follows by mere induction.

\section{Trace preservation of $\Lambda(t)$}

As discussed in the main text, the integrand in MT's \eq(10) is a CPT map and thus, once applied to $\rho_0$, it yields a state having unitary trace (clearly, ${\rm Tr}\rho_0\ug1$). Hence,
 \begin{eqnarray} \label{trace}
{\rm Tr}\left\{\Lambda(t)[\rho_0]\right\}&\ug&\sum_{k=1}^\infty\Gamma^{k\!-\!1}\,{\rm Tr}\left\{e^{-\Gamma t}\int_0^t \!\!\!dt_1\!\!\!\int_0^{t_1}\!\!\!\!dt_2\!\cdot\!\cdot\!\cdot\!\!\int_0^{t_{k\meno2}}\!\!d t_{k\meno1} \right\}\nonumber\\
&\ug&e^{-\Gamma t}\sum_{k=1}^\infty\Gamma^{k\!-\!1}\frac{t^{k\meno1}}{(k\meno1)!}\ug  e^{-\Gamma t}e^{\Gamma t}\ug1\,\,.
\end{eqnarray}

\end{document}